\documentclass[14pt]{revtex4}

\usepackage{graphicx,epstopdf}

\begin{document}

\title{Answer to the Comments on the paper ``A study of phantom scalar field cosmology using Lie and Noether symmetries''  by A. Paliathanasis, et al. arXiv:1605.05147 }

\author{Sourav Dutta\footnote {sduttaju@gmail.com}}
\author{Subenoy Chakraborty\footnote {schakraborty.math@gmail.com}}
\affiliation{Department of Mathematics, Jadavpur University, Kolkata-700032, West Bengal, India.}

\begin{abstract}
	
	This is a short communication in reference to the comments in (arXiv: 1605.05147) of our paper in Int. J. Mod. Phys. D 25, 1650051 (2016). We have tried to answer to the comments or criticism in the paper.\\
	
	
\end{abstract}

\maketitle

	 Recently, a short paper has been put on (arXiv: 1605.05147) by Paliathanasis et al.[1] and possibly will be published in Int. J. Mod. Phys. D . They have criticized our paper[2] in Int. J. Mod. Phys. D 25, 1650051 (2016) on phantom cosmology stating that the results of our paper are not at all new and can be obtained more easily by a co-ordinate transformation. We essentially disagree with most of their comments and maintain that our paper contains important results which are new and not have been published earlier and that these authors' Comments are essentially not relevant to our work. \\
	 
	 We find that unfortunately these authors have missed the spirit of our work. They have claimed that for the equation (3) of their paper  our contention that the Lie point symmetries or Noether symmetries give constraints which determine the unknown quantities is not true. This statement is not correct.They have mentioned the well known result that the admitted group of invariant transformations of dynamical system is independent of the co-ordinate system. The authors of this note are completely silent about the Lie-point symmetries and the corresponding solutions reported by us..Also they have not uttered a single word about the Noether symmetry results given in equations (46) and (47). Are these Noether symmetries available in the literature as they have mentioned ? We do not find this to be true. Ref. 2 of their paper deals with Noether symmetry with the Lagrangian after the transformation and their Lie symmetry is related to the symmetries reported in our  eqs. (40)- (42).(Note: Pl. make sure that the above couple of sentences are not contradicting each other). Also their transformation cannot give the Lie and Noether symmetries associated with our equations.(46)-(47). Further their references 3-5 deal with general scalar field cosmology and have nothing to do with symmetries. Also their final statements regarding the original works of Lie and Noether, which we highly respect, have as far as we can see in no way prohibit our results.\\
	 
	 In short we feel that the Comment of Paliathanasis et al. written in a hasty manner and their contention in no way reduces the significance of our work.\\

\end{document}